**Thoughts on Learning Human and Programming Languages**

Daniel S. Katz and Jeffrey C. Carver

**Abstract**:

This is a virtual dialog between Jeffrey C. Carver and Daniel S. Katz on how people learn programming languages. It's based on a talk Jeff gave at the first US-RSE Conference (US-RSE'23), which led Dan to think about human languages versus computer languages. Dan discussed this with Jeff at the conference, and this discussion continued asynchronous, with this column being a record of the discussion.

**Introduction**:

At the first US-RSE Conference (USRSE23) in October in Chicago, in his presentation on "Peer Code Review in Research Software," Jeff Carver discussed how people learn programming languages. Specifically, he stated:

*Generally people learn new languages by reading the language before writing the language themselves. However, for learning programming languages, we often do the opposite. We ask people to write a language before they learn how to read it.*

This led me to think about human languages vs. computer language, and a short discussion with Jeff at the conference led to this blog post where we each provide some initial thoughts and then discuss them together.

**Dan's thoughts:**

Human languages are generally used to communicate ideas from one person to at least one other person. These languages were initially generally spoken - expressed through sound (or gestures, in the case of signed languages), and similarly interpreted orally (or visually for signed languages).

Historically, it became important to have a way to record language, leading to writing to express it and reading to interpret it. In fact, for quite a while, writing was the one place where a person would communicate something to themself in the future (notes), though this has more recently been augmented by recorded speech as notes.

And even more recently, automated systems can translate between written and spoken language, with good and improving accuracy.

On the other hand, computer languages were initially intended for humans to express instructions in writing for computers to interpret and act upon.

All these languages can be also used to provide detailed instructions for people about how something should be done. One can think of a spectrum of instructions, from normal narrative text in a human language to formal mathematics to pseudocode to a computer language intended for both human and computer usage (e.g., Python, R, C, Fortran) to a computer language intended for computer usage (e.g., assembly). This set of options also generally moves from fairly loose to more detailed specification of activities, as well from high-level to low-level instructions.

In the remainder of this post, I want to consider how these languages are learned, thinking about writing, speaking, and reading.

Children learn human languages by both listening and speaking initially, based on interaction with others (conversation). They then later learn reading and writing, often in a more formal and structured setting.

On the other hand, adults learning a new human language often start with reading and writing, then learn speaking and listening, or do all four simultaneously. Perhaps this is because speaking and listening are more inherent, but once reading and writing have been learned for one language, they can be used in the process of learning a new language. While speaking and listening involve interaction and feedback, where the feedback in reading and writing happens is much less clear, and they can be done, at least in part, by oneself. At some level, the feedback in reading might be based on if it leads to comprehension, and the feedback in writing could be in comparison to other "correct" writing.

Adults learning computer languages is different, as these languages aren't really spoken or heard, but are primarily read and written. However, writing computer languages is somewhat like speaking human languages, as this is where feedback is generated, by the computer in the case of computer languages and the listener in the case of human languages.

Children commonly learn programming graphically, which allows them to start learning earlier than if they had to use text. In this case, the children are learning a form of symbolic instruction. However, they are not learning the same language as adults do.

While success in speaking a human language can generally be seen by how a human listener responds, typically in the same language, success in writing a computer language can be seen in how the computer acts, as the computer is a system that acts on instructions, not one that thinks independently. (Even systems using large language models and generative AI are not based on thinking, but are based on acting on instructions.)

In general, this feedback and interaction seems very important in learning any language, whether human or computer, or even another skill. Peter Norvig in "[Teach Yourself Programming in Ten Years](#)" says "*Play.* Which way would you rather learn to play the piano: the normal, interactive way, in which you hear each note as soon as you hit a key, or "batch" mode, in which you only hear the notes after you finish a whole song? Clearly, interactive mode makes learning easier for the piano, and also for programming. Insist on a language with an interactive mode and use it."

To summarize, this view of learning both human and computer languages is based on two parties and direct iteration as much as possible, though a single party may be able to learn reading and writing a language based on using a corpus as a stand-in for the second party.

When we consider peer review and pair programming for computer languages, we add a third party. In peer review, one person tries to understand a computer program written by another person, using a computer as the third party. This is also related to debugging. In pair programming two people work together to write computer code, again using the computer as the third party.

**Jeff's Response:**

Before making some conclusions on the similarities between learning a human language and learning a programming language, I will first describe some general observations about the similarities and differences between these two types of languages. While there are many similarities, programming languages differ from traditional human languages in a number of ways. These differences impact how one learns and becomes proficient in a language.

First, unlike human languages, which focus primarily on human-human communication, programming languages focus on both human-human communication and human-machine communication, at least for higher-level programming languages. This dual purpose adds some additional constraints on what "effective communication" means and results in some differences in our thinking about how best to learn the languages.

Second, effective communication includes both syntactic and semantic elements. However, the way these concepts appear differs across types of languages. For syntax, in computer languages the syntax requirements are quite strict. We have tools that can verify, and enforce, correctness, i.e., the compiler. If the syntax is not correct, then the receptor of the communication, the machine, will not be able to understand the communication. Conversely, for human languages, while there are syntactic rules, which can also be checked by tools, e.g., Grammarly, the receptor of the communication, a human, can still understand communication that is syntactically flawed. For semantics, in computer languages, we cannot separate the semantics from the syntax. In other words, the receptor of the language, the computer, does not have to interpret what the author meant. It just executes the commands provided. The author may make semantic mistakes in their writing, but the computer plays no role in this type of mistake. Conversely, for human languages, correct syntax does not automatically result in correct semantic understanding by the receiver. Text that is syntactically correct may be understood to have a different meaning than the original author intended. Of course, the closeness of the correspondence between syntax and semantics varies by language, so the statements above are generalizations.

Third, as mentioned above by Dan, the type and concreteness of the feedback the receptor provides the communicator differs between human languages and programming languages. Let's assume that the communicator knows exactly what they are attempting to communicate. In the context of a programming language, the feedback from either the compiler (for syntax) or the run-time system (for semantics) will tell the communicator whether their language had the intended outcome. Further, if the outcome is not what the communicator intended, the type of feedback (either compiler errors or unexpected output from the run-time system) will give the communicator concrete information about how their language was incorrect. Conversely, in the context of a human language, it may not always be clear to the communicator whether they were properly understood. In cases where there is a misunderstanding, the feedback will often be less clear about what is incorrect.

Finally, in both human and programming languages there are families of languages that have features in common. Often one language derives from the other. For example, in human languages there is the family of Romance languages derived from Latin, which include French, Spanish, Portuguese, and Italian. In the context of programming languages, there are both types of languages (e.g., declarative or imperative) and languages that derive from others (e.g., C++ derives in part from C). The benefit of language families is that knowing one language in the family makes it easier to learn another language within the same family.

Building on the thoughts above, I will now turn to some observations about learning new languages.

First, to learn a new language, of any kind, a person needs at least two items: the rules of the language and a model of what constitutes a good use of the language. For both types of languages, there is a list of syntactic rules describing the types of constructs that are allowed and those that are not allowed. However, the model of the language for human languages can be either spoken or written, while the model for programming languages can only be written. In either case, there are two ways the language could be learned. The learner could either study and learn all the syntactic and semantic rules for the language and translate those to written text, or the learner could read through the models that provide examples of how to communicate in that language. In reality, there is a need for both of these approaches. While one must learn the rules, those rules can best be understood by seeing them implemented in a model of the language. To illustrate this concept in another domain, one could learn to drive a car simply by studying the traffic rules and the owner's manual for their vehicle. However, you would likely not want to ride with a person if they had never seen examples, good and bad, of other drivers. Those illustrations matched together with the theoretical knowledge from the traffic rules and the owner's manual provide the best way to learn how to drive a car. Therefore, for both types of languages, there is a need to have good examples or models. For programming languages, these models exist only in writing (not spoken), so one must learn how to read the language.

Second, because the semantics and syntax of programming languages are so closely linked, the written models provide good concrete illustrations of how to communicate different concepts. I argue that, while there are many applications of the concepts, there are relatively few fundamental concepts that are communicated with programming languages. Conversely, in human language, the possibilities are endless. Therefore, these written models of language serve as a good way for someone to learn these fundamental concepts in programming languages.

**Final Discussion**

Dan: Thinking about your comment that computers can't understand communication that is syntactically flawed while humans can reminds me of something I read recently , saying that this property varied across human languages, and theorizing that part of the reason for English's popularity was that it was particularly good at or accepting of this (see second and third letters in [https://www.theguardian.com/science/2023/dec/29/the-costs-and-benefits-of-english-as-a-lingua-franca](https://www.theguardian.com/science/2023/dec/29/the-costs-and-benefits-of-english-as-a-lingua-franca)). On the other hand, many argue that English's popularity is due to historical and power factors. Maybe it's some of both.

Dan: Your comment that for "a human language, it may not always be clear to the communicator whether they were properly understood. In cases where there is a misunderstanding, the feedback will often be less clear about what is incorrect." reminds me of [active listening](), where part of the goal is to improve human communication and understanding.

Jeff: Could we consider that a compiler/interpreter is really the original form of active listening? Maybe human language trailed behind software languages in terms of reinforcing understanding.

Dan: Finally, in regards to your comments about learning to drive, I'm reminded that US driving exams typically have both a written and driving part. Driver's education classes used also have both parts. In both cases, the written part comes first and is necessary to pass, but isn't sufficient without also passing/learning the driving part. I wonder if this order matches the order of learning languages? Probably not, as babies learn to speak only from practice without written instructions. Could one learn a computer language only by examples and trial-and-error?

Jeff: I think this point is consistent with my point about needing to know how to read programming languages. Even if babies learn to speak from practice and trial-and-error, they still need a model (e.g. they are mimicking the more mature version of the language they hear spoken by more mature people around them). To learn programming, new developers (i.e. babies) need to practice by mimicking the more mature version of the language written by more mature developers.

Dan & Jeff: We hope this virtual discussion is useful for readers who are thinking about these issues, particularly in the context of training for both those who want to make a career in work in research software, and those for whom research software can be a key element that enables their research. We welcome further discussion via the US-RSE Slack's #education-training channel. (To get access to the US-RSE Slack, join US-RSE at no cost at [https://us-rse.org/join/](https://us-rse.org/join/); US-RSE membership is open to all who are interested in research software and the people who develop and maintain it.)